# Current Driven Domain Wall Depinning in Notched Permalloy Nanowires

Candra Kurniawan[1, a)] and Dede Djuhana[2, b)]

[1]*Research Center for Physics, Lembaga Ilmu Pengetahuan Indonesia, Tangerang Selatan, Indonesia*
[2]*Departement of Physics, Universitas Indonesia, Depok, Indonesia*

[a)]cand002@lipi.go.id
[b)]dede.djuhana@sci.ui.ac.id

**Abstract.** In this work, we have investigated the domain wall (DW) depinning behavior in the notched nanowire by a micromagnetic simulation. A transverse domain wall (TW) was initially positioned at the center of notch and 1 ns length current pulse was applied to depin the DW with respect to the notch size $s$ and the wire width variation. We have observed the depinning current density $J_d$ which was a minimum current to escape DW from the notch. It was found that the depinning current density decreased as the wire width and the notch size increased. In the depinning process, we observed the inner structure of DW generally transformed from TW to anti-vortex wall (AVW). Interestingly, for the case of $s$ less than 70 nm, AVW formed and depinned closely to the period when current pulse was active, while for $s$ larger than 70 nm, AVW formed until the current pulse went to zero and then depinned after flipped TW was formed. It can be explained that the transformation of DW inner structures were affected by the spin torque energy and contributed to DW depinning behavior from the notched nanowires.

## INTRODUCTION

Recently, manipulating and controlling the motion of magnetic domain wall (DW) has been an interesting area in the spintronics research. Both experimental and theoretical aspects of the DW behavior in the nanowire have been developed in relation with the application of spintronics devices such as magnetic logics [1–3] and magnetic memory [4–6]. One of the promising application based on the DW motion in the nanowire is the racetrack memory (RM) proposed by Parkin et al. [5,7]. The RM used DW motion for storing data which is injected by spin polarized current. This kind of data storage presumes to perform approximately 100 times faster than commercial technology such as flash or harddisk drive (HDD) memory [7]. Many studies reported about the depinning process driven by magnetic field [8–11] and spin polarized current [5,12–15] in the notched nanowires. It was known that the depinning process was directly influenced by the shape of the constriction as the pinning site and the DW inner structure. However, the depinning process in a symmetrical triangular notch driven by spin polarized current corresponding to the notch size variation and the DW transformation during the depinning process is not well understood recently.

In this work, we have investigated the DW depinning behavior in the notched ferromagnetic nanowires driven by spin polarized current by means of a micromagnetic simulation. The depinning current to push DW out from the notch was affected by the notch size and wire width. A transformation of DW inner structure occurred from transverse wall to anti-vortex wall during the depinning process. It also explained that the spin torque energy contributed to DW transformation.

## COMPUTATIONAL METHODS

The nanowire model was using a long and thin rectangular shape. The length and thickness of the wires were 2000 nm and 5 nm respectively. The wire widths were 150 and 200 nm. The double symmetrical triangular notch was introduced in the middle of the wires as shown in Fig. 1. The notch size ($s$) was varied from 10 nm to 200 nm with respect to the notch center. The material used was Permalloy which has material properties of saturation magnetization $M_s = 8 \times 10^5$ A/m, the exchange stiffness constant $A = 13 \times 10^{-12}$ J/m and zero magnetocrystalline anisotropy.

The simulation was performed using public free micromagnetic solver, OOMMF package [16]. The extension of spin transfer torque (STT) made by IBM was applied to include the interaction of local magnetization and electrical current based on Eq. (1) as proposed by Zhang and Li [17]. The simulation was using an adiabatic process of spin transfer torque as idealization following the Landau-Lifshits-Gilbert (LLG) equation,

$$\frac{d\vec{m}}{dt} = -\gamma \vec{m} \times \vec{H}_{eff} + \alpha \vec{m} \times \frac{d\vec{m}}{dt} - (\vec{u} \cdot \nabla)\vec{m} \tag{1}$$

where $\gamma$ is the gyromagnetic ratio of an electron and $\alpha$ is Gilbert damping constant which was set to 0.01 [11,18]. The $\vec{m}$ is vector along local magnetization and $\vec{H}_{eff}$ is vector of micromagnetic effective field. The electrical current was included in the $\vec{u}$, which is a velocity directed along the direction of electron motion (spin current) that proportional to the amplitude of the torque: $u = JPg\mu_B/(2eM_s)$, where $J$ is the current density, $P$ is the spin polarization, $g$ is the Landé g-factor, $\mu_B$ is the Bohr magneton, $e$ is the electron charge, and $M_s$ is the spontaneous magnetization of the ferromagnetic layer. The cell size was $5\times5\times5$ nm$^3$ with respect to the exchange length of Permalloy which about 5 nm [19].

The transverse typed DW was initiated to be positioned at the center of notch with the head-to-head spin structure. The current pulse of 1 ns length was injected to the wire at the +x-axis (electron) direction with raise and drop time of 0.01 ns as illustrated in Fig. 1. Then the current pulse was turned off and the domain wall was moved under zero current condition. In this simulation, the magnetization in the nanowire was only triggered by electrical current without external magnetic field. The DW depinning current ($J_d$) was investigated by modifying the magnitude of input current density ($J$). The depinning condition was determined by the latest domain wall boundary that moves out from the edge of the notch.

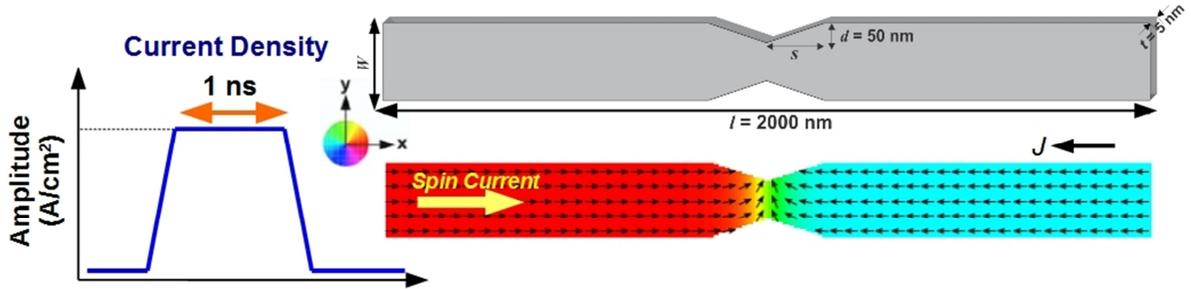

**FIGURE 1.** Dimension and geometry of the symmetrical notch Permalloy nanowires with initial head-to-head spin structure at the center of the notch. Injected spin current with 1 ns length electric current pulse is applied in the +x direction, and the pulse amplitude represents the current strength. The color disk represents spin configuration for each pixel.

## RESULTS AND DISCUSSIONS

In our study, we have observed that DW can be triggered by electrical current only. When the current is flowing through the nanowire, the torque on the DW was raised and the transfer of momentum from the conduction electron to valence electron of the material occurred. We observed that the DW could escape from the notch if the current density is large enough to surpassed the depinning current density ($J_d$).

The results of DW motion induced by electrical current are displayed in Fig. 2 for $W = 150$ nm and $s = 50$ nm. The figure shows the evolution of DW dynamics at current density values for below ($J < J_d$) and proportional to depinning current density ($J = J_d$) as the time increases. It also shows the DW behavior during the pulse current injection ($t = 1$ ns) and after the current was removed. Figure 2(a) shows that for injected current below $J_d$, the DW oscillates around the notch area. The transverse DW moved forward along the notch and later moved backward after it reached the notch edges. The DW position remains stable after the injected current was removed and pinned back in the notch center. It can be interpreted that below the depinning point, the DW cannot escape from the notch because of the notch that acts as a pinned potential.

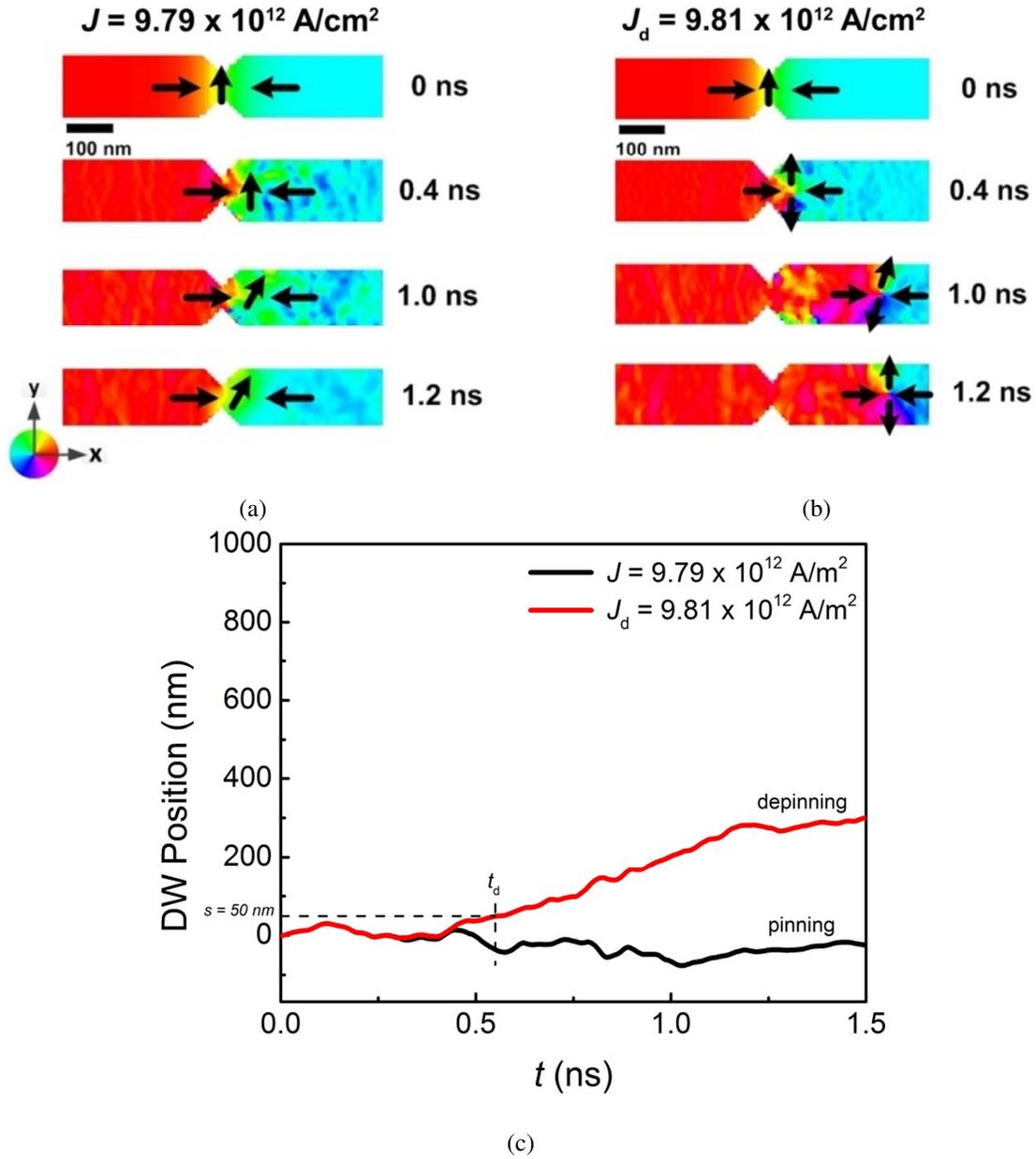

**FIGURE 2**. Evolution of DW structures in the nanowire with $W$ = 150 nm and $s$ = 50 nm at injected current (a) below ($J < J_d$) and (b) proportional to depinning current density ($J = J_d$). The arrows represents spin configuration formed around DW as the time increases. (c) The DW position comparison during pinning and depinning condition is plotted and shows the propagation dynamics of DW. The time when DW leaves the notch area is noted as the depinning time ($t_d$).

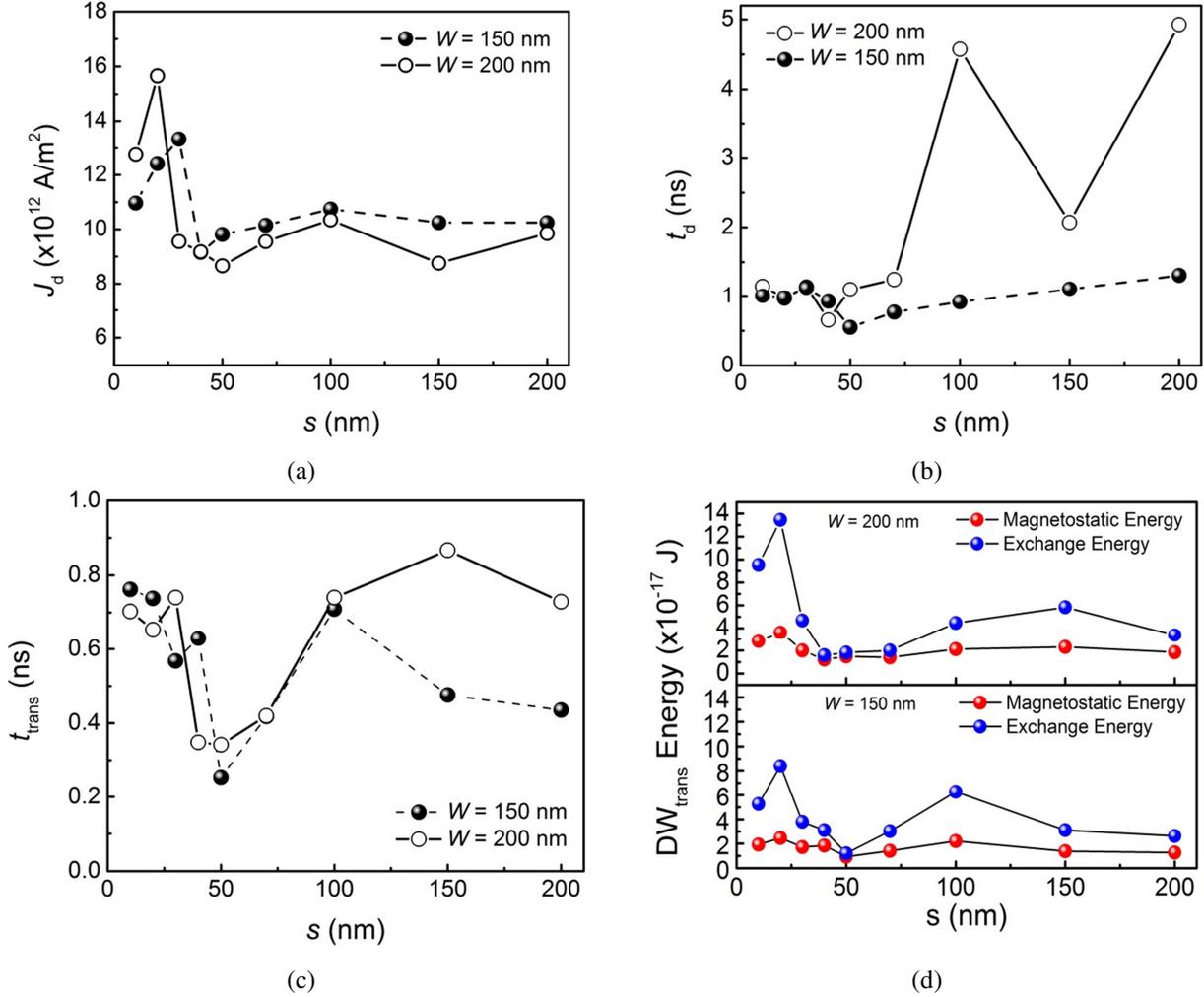

**FIGURE 3**. The DW depinning behavior with respect to the notch size variation, (a) the depinning current density, $J_d$, (b) the depinning time, $t_d$, and (c) the DW structure transition time, $t_{trans}$, of 150 nm (black circle) and 200 nm (white circle) wire widths. The graphic (d) shows the spin torque energy profile at the transition time with respect to the notch size.

The DW behavior at $J = J_d$ was quite different as shown in Fig. 2(b). It can be seen that the induced current density was large enough to move out the DW from the notch area. During this depinning process, the DW structure was changed from transverse wall into anti-vortex wall. The anti-vortex core was raised from the notch bottom then moved along the DW and escaped from the notch. For example, As shown in the Fig. 2(b), the $J_d$ for wire width of 150 nm and $s = 50$ nm was about $9.81 \times 10^{12}$ A/m$^2$. The value is analogues with depinning spin drift velocity ($u_d$) values as the input of simulation which was about 493 m/s.

To understand the change of DW position during current pulse injection and after the current was removed, we plotted the approximated DW position as the increasing of observed time as shown in Fig. 2(c). The DW position was calculated by using the ratio of average magnetization on x-axis direction over the saturated magnetization value, $\bar{m}_x = M_x/M_s$. The head-to-head DW as the initial condition was defined to be located at 0 nm (center of the notch). As DW propagates through the wire on +x direction, the value of $\bar{m}_x$ was increases to the positive value. The ratio then compared to the total distance of DW track which is half of wire length ($0.5l = 1000$ nm) to get the approximated DW position at x-axis direction. Figure 2(c) shows that the DW oscillates around the notch area in both below and at the depinning point. The DW oscillation amplitudes are around the notch length ($s$) of 50 nm. For $J < J_d$, the DW oscillation was damped during the current pulse injection period. At the depinning point ($J = J_d$), the same oscillation behavior also observed before the DW escaped from the notch area. This behavior is explained as the effect of the DW inner structure changing during depinning process.

Furthermore, the effect of notch size ($s$) to DW depinning behavior was investigated. The effect of notch size was studied by determining $J_d$ values as $s$ increases. The $J_d$ values for all $s$ variation are depicted in Fig. 3(a). In the smaller $s$, the $J_d$ values for wire width of 200 nm are larger than of 150 nm. The highest value of depinning current density of $15.6 \times 10^{12}$ A/m$^2$ was observed in the notch size and wire width of 20 nm and 200 nm, respectively. The depinning current at the nearly same size is comparable with the work of Huang and Lai [20] that achieved at $17.5 \times 10^{12}$ A/m$^2$ for $s$ of 17.5 nm. The depinning current density are sharply reduced for $s > 30$ nm on both wire widths. Figure 3(a) indicates that depinning current density ($J_d$) was decreased as the increasing of wire width. This result is in good agreement with the work of Djuhana et al. [11] for the case of DW depinning by pulse magnetic field only. The Fig. 3(a) also shows that the depinning current density for $s > 30$ nm are relatively insensitive with the increasing of notch size. The standard deviation of $J_d$ for $s > 30$ are 5.3% and 7% for wire width of 150 nm and 200 nm, respectively. It means that the significant effects of notch size were observed only for smaller notch size ($s \leq 30$ nm).

Another aspect of DW depinning behavior was the depinning time. The depinning time ($t_d$) was an important parameter that directly correlated to the speed of information transfer. If the DW escapes faster from the pinning site, it could be translated that the data moved as quick as the DW displacement. The profile of depinning time as the increasing of notch size is provided in Fig. 3(b). In the figure, the depinning time is less sensitive to the wire widths for the smaller $s$. For $s \leq 70$ nm, the difference of $t_d$ values for both wire widths were not significant. The biggest and the smallest difference of depinning time at this range were 0.55 ns and 0.01 ns at notch size of 50 nm and 20 nm, respectively. Compared to the $t_d$ values at range of $s > 70$ nm, there are huge differences of depinning time at the order of nanoseconds. Thus, the wire width variation only affects depinning time at larger notch size ($s > 70$ nm). For adiabatic torque, the depinning time at the wire width of 150 nm was less sensitive compared to the larger one ($W = 200$ nm).

Previous studies of DW depinning has shown that the DW inner structures were changed from transverse wall (TW) to anti-vortex wall (AVW) during the depinning process [20–22]. Accordingly, we have analyzed the DW structure transition time ($t_{trans}$) that described the time when AVW core raised in the notch area when DW change its structures from TW to AVW. The profile of $t_{trans}$ as the increased of notch size ($s$) is showed in Fig. 3(c). There is a minimum range of $t_{trans}$ values from $s = 40$ nm to $s = 70$ nm. The decreasing of $t_{trans}$ in that range is related with the difference of exchange and magnetostatic energy as showed in Fig. 3(d). The starting time of DW structure change is faster if the difference of spin torque energy become smaller. In Figure 3(c), we also shows that the $t_{trans}$ values did not changed significantly by varying wire width, except for $s > 100$ nm. Hence, the wire width effect can be neglected for transition time analysis at $s \leq 100$ nm.

Based on the results above, we suggest that the smaller wires are more capable to maintain the data rate because the small difference of notch size will not affect to DW dynamics. Moreover, nanowire with wire width and notch size of 150 nm and 50 nm is recommended because of the relatively small depinning current ($J_d$), DW transition time ($t_{trans}$), and faster depinning time ($t_d$), especially for adiabatic process.

## CONCLUSIONS

We have studied the micromagnetic simulation of domain wall (DW) depinning in the Permalloy nanowires. The domain wall dynamics was triggered using pulse electrical current and the constrictions was introduced in the nanowires geometries. It is known that the constrictions such a notch can serve as pinning potential for DW in which it would require certain amount of electrical current for escaping DW from the notch called as depinning current density $J_d$. In the depinning process, we observed the inner structure of DW was generally transformed from TW to anti-vortex wall (AVW). Interestingly, for the case of s less than 70 nm, AVW formed and depinned closely to the period when current pulse was active, while for s larger than 70 nm, AVW formed until the current pulse went to zero and then depinned after flipped TW was formed. The characteristic of DW structure transition was contributed to DW depinning behavior such as depinning and transition time. It can be explained that the transformation of DW inner structures were affected by the spin torque energy during the depinning process.

## ACKNOWLEDGMENTS

This work was supported by program Riset Mandiri 2015 from Research Center for Physics, Lembaga Ilmu Pengetahuan Indonesia (LIPI) and hibah PUPT 2015 from Kementerian Riset, Teknologi, dan Pendidikan Tinggi Republik Indonesia No. 0527/UN2.R12/HKP.05.00/2015.